\documentclass[aps,prl,showpacs,floatfix,twocolumn,amsmath,amssymb,preprintnumbers]{revtex4-1}
\usepackage{mathrsfs}
\usepackage[figuresright]{rotating}
\usepackage{amsmath}
\usepackage{amssymb}
\usepackage{siunitx}
\usepackage{graphicx}
\usepackage{color}
\usepackage{dcolumn}
\usepackage{bm}
\usepackage[breaklinks=true,colorlinks=true,linkcolor=blue,urlcolor=blue,citecolor=blue]{hyperref}
\UseRawInputEncoding   

\usepackage{longtable}

\usepackage{multirow}
\usepackage{setspace}

\usepackage{soul}

\makeatletter

\newcommand{\Rmnum}[1]{\expandafter\@slowromancap\romannumeral #1@}
\makeatother

\begin{document}

\title{Elastic properties of Cu-6wt\%Ag alloy wires for pulsed magnets investigated by ultrasonic techniques}

\author{Ziyu Li$^{1,2}$} 
\author{Tianyi Gu$^{1,3}$} 
\author{Wenqi Wei$^{1,3}$} 
\author{Yang Yuan$^{1,2}$} 
\author{Zhuo Wang$^{1,2}$} 
\author{Kangjian Luo$^{1,2}$} 
\author{Yupeng Pan$^{1,2}$} 
\author{Jianfeng Xie$^{1,3}$} 
\author{Shaozhe Zhang$^{1,3}$} 
\author{Tao Peng$^{1,3}$} 
\author{Lin Liu$^{4}$} 
\author{Qi Chen$^{1,3}$} 
\email[]{qichen@hust.edu.cn}
\author{Xiaotao Han$^{1,3}$} 
\email[]{xthan@mail.hust.edu.cn}
\author{Yongkang Luo$^{1,2}$} 
\email[]{mpzslyk@gmail.com}
\author{Liang Li$^{1,3}$} 
\address{$^1$Wuhan National High Magnetic Field Center, Huazhong University of Science and Technology, Wuhan 430074, China;}
\address{$^2$School of Physics, Huazhong University of Science and Technology, Wuhan 430074, China;}
\address{$^3$State Key Laboratory of Advanced Electromagnetic Engineering and Technology, Huazhong University of Science and Technology, Wuhan 430074, China;}
\address{$^4$School of Materials Science and Engineering, State Key Lab for Materials Processing and Die $\&$ Mold Technology, Huazhong University of Science and Technology, Wuhan 430074, China}

\date{\today}

\begin{abstract}

Conductor materials with good mechanical performance as well as high electrical- and thermal-conductivities are particularly important to break through the current bottle-neck limit ($\sim 100$ T) of pulsed magnets. Here we perform systematic studies on the elastic properties of the Cu-6wt\%Ag alloy wires, a promising candidate material for the new-generation pulsed magnets, by employing two independent ultrasonic techniques - resonant ultrasound spectroscopy (RUS) and ultrasound pulse-echo experiments. Our RUS measurements manifest that the elastic properties of the Cu-6wt\%Ag alloy wires can be improved by an electroplastic drawing procedure as compared with the conventional cold drawing. We also take this chance to test the availability of our newly-built ultrasound pulse-echo facility at Wuhan National High Magnetic Field Center (WHMFC, China), and the results suggest that the elastic performance of the electroplastically-drawn Cu-6wt\%Ag alloy wire remains excellent without anomalous softening under extreme conditions, e.g., ultra-high magnetic field up to 50 T, nitrogen / helium cryogenic liquids.


\end{abstract}


\maketitle

\section{\Rmnum{1}. Introduction}

High magnetic field plays an important role in modern material science and condensed matter physics. On the one hand, magnetic field is one of the key control parameters, and many intriguing phenomena can be induced by high magnetic field, such as integer / fractional quantum Hall effect \cite{klitzing1980new,TsuiD-PRL1982,LaughlinR-PRL1983}, quantum magnetization plateau \cite{Takigawa-SrCu2B2O6PRL2013}, quantum critical point \cite{JiaoL-PNAS2015}, Fulde-Ferrell-Larkin-Ovchinnikov (FFLO) state \cite{FuldeP-PR1964,LarkinA-SovPhysJETP1965}, reentrance of superconductivity \cite{RanS-NP2019}, etc. On the other hand, high magnetic field also acts as a ``magnifier" to physical ensembles, e.g., by reducing the magnetic length $l_B\equiv\sqrt{\hbar/eB}$, quantum oscillations are amplified, thus Fermi surface topology of materials can be more resolvable \cite{Ramshaw-YBCOSdH}; by enhancing the ratio $B/T$, the thermal fluctuations will be suppressed so that phenomena related to quantum fluctuations can be more observable; by promoting the signal-to-noise ratio (S/N $\propto B^{3/2}$), the resolution of nuclear magnetic resonance can be greatly lifted \cite{Levitt-NMR,LiuQ-PFNMR}; etc. In this sense, pulsed field, currently the only non-destructive method to create magnetic field larger than 46 T, has attracted great interest in the past decades. However, due to multiple factors including electrical conductivity, thermal conductivity, strength of conducting wires, electromagnetic coil winding technique, and others, so far the amplitude of pulsed field has been limited to $\sim 100.75$ T (world record by Los Alamos National Lab., USA) \cite{Nguyen-LANL100.75T}.

\begin{figure*}[htbp]
\vspace*{-0pt}
\hspace*{-0pt}
\includegraphics[width=18cm]{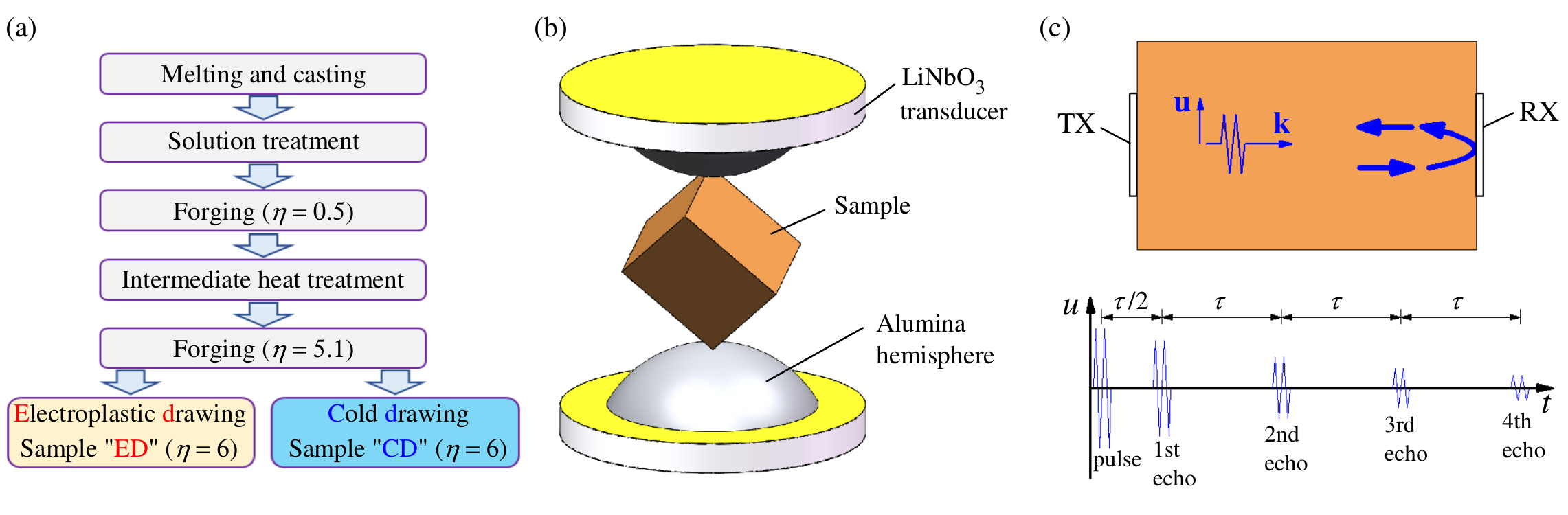}
\vspace*{-15pt}
\caption{(a) Flow chart for Cu-6wt\%Ag alloy wires growth; the two samples labeled by ``ED" and ``CD" were prepared in electroplastic drawing and conventional cold drawing, respectively. (b) Scheme of RUS experimental set-up. (c) A sketch of ultrasound pulse-echo measurements. }
\label{Fig1}
\end{figure*}

One of the central issues to develop ultra-high pulsed-field magnet is to manufacture proper conducting wires with excellent mechanical performance. Cu-Nb macro-composite and Cu-Ag alloy are the two materials commonly used for inner coil of pulsed magnets. For instance, in 1990s, Japan successfully created a high magnetic field exceeding 80 T by Cu-Ag alloy \cite{14sakai1997ultra}; more recently, Wuhan National High Magnetic Field Center (WHMFC) in China has created a magnetic field $\sim$ 94.8 T by using Cu-Nb conductors \cite{14han2017pulsed,Xie-WHMFC2023}. Compared with Cu-Nb macro-composite, the conductivity of Cu-Ag alloys is generally higher, which is beneficial to reducing Joule heating; and moreover, to reach a similar strength, the draw ratio $\eta$ required is relatively smaller \cite{14sakai1997ultra}. For Cu-Ag alloy, the mechanical and electrical properties depend strongly on the fabrication methods. For example, Y. Sakai \textit{et al} reported ultra-high strength and high conductivity in Cu-Ag cold drawing combined with intermediate heat treatment \cite{Sakai-APL1991,Sakai-IEEETransMagn1994, Sakai-ActaMetallMater1995}, in particular, the optimized Cu-24wt\%Ag wire with $\eta=5.8$ shows ultimate tensile strength (UTS) of 1.5 GPa and an electrical conductivity of 65\% IACS (International Annealed Copper Standard) at room temperature \cite{14sakai1997ultra}. For Cu-6wt\%Ag with lower expense, the electrical conductivity can be enhanced to 80\% IACS, nevertheless, the UTS is much lower \cite{cuag6sakai2016development}.
This trade-off relation between strength and electrical conductivity of Cu¨CAg alloys restricts their scope of application \cite{KongL-CuAg2024}.
Recently, Chen \textit{et al} successfully manufactured the Cu-6wt\%Ag alloy wire of rectangular section $3\times5$ mm$^2$ via electrically-assisted drawing. It was found that by an electroplastic
drawing (ED) process, the UTS, elongation and electrical conductivity of Cu-6wt\%Ag alloy wire can be enhanced by 8\%, 4\% and 30\%, respectively \cite{ZhangL-CuAgThesis,Chen-CuAgUnpublish}, strongly suggesting the potential application for the 110 T project at WHMFC.

However, we noticed that although the fabrication method and the resultant UTS of Cu-Ag alloys have been well studied, little has been known about their elastic properties, especially at liquid-nitrogen temperature where these conducting wires typically work for pulsed magnets. In addition, we are also curious about their performance under ultra-high field, which, in principle, should have been a prerequisite for their use as conductor materials in pulsed magnets, but was omitted here before. In this paper, by a combination of resonant ultrasound spectroscopy (RUS) and ultrasound pulse-echo experiments, we report the elastic properties of Cu-6wt\%Ag alloy wires at both atmosphere and extreme conditions. The main results are summarized in Tab.~\ref{Tab.1}. Our results show that the elastic properties of Cu-6wt\%Ag alloy wires can be improved by an electroplastic drawing procedure, and the performance remains excellent without anomalous softening under ultra-high magnetic field up to 50 T and low temperature down to 1.9 K.

\section{\Rmnum{2}. Experimental details}

Oxide-free copper and silver with a purity of 99.99\% (Cu-6wt\%Ag) were selected and melted in an argon atmosphere. The Cu-Ag alloy ingot of 120 mm in length and 90 mm in diameter was then prepared by conventional cast method. The ingot was heated at 780 $^\circ$C for 2 hours and quenched in water for solid solution treatment. Subsequently, the ingot was forged to a diameter of approximately 70 mm and underwent heat treatment at 450 $^\circ$C for 20 hours, followed by cooling to room temperature in the furnace. The aged ingot was hot forged to 40 mm in diameter. Finally, the rod with a diameter of 12 mm was produced by rotary forging and drawn into a rectangular conductor wire of 3$\times$5 mm$^2$. Drawing reduction was given in teams of logarithmic strain by draw ratio $\eta\equiv\ln{(A_0/A)}$, where $A_0$ and $A$ represent the initial and final cross-sectional areas, respectively. When the draw ratio reached $\eta=5.1$, a part of the alloy was drawn by electrically assisted drawing (sample ED), while the other part was drawn by conventional cold drawing (sample CD). The deformation process is described in Fig.~\ref{Fig1}(a). In this work, the electrically assisted drawing processing instrument was used by adding a pulse power supply to a conventional wire-drawing machine. A pulse square-wave electric source was self-made to provide a high-density current during the drawing. 
Electrically assisted drawing applies current to the deformation area of the metal wire. Current density of 50 A/mm$^2$, duration time of 10 ms and duty ratio of 50 \% were applied in the experiment. Under the condition of a drawing speed of 5 m/min, a tungsten steel alloy drawing die was used. In order to compare the effect of drawing process on the mechanical properties of the material, both samples were taken when the draw ratio $\eta=6$. More details about the sample preparation can be found in Ref.~\cite{ZhangL-CuAgThesis,Chen-CuAgUnpublish}.

\begin{figure}[htbp]
\vspace{-0pt}
\hspace{-0pt}
\includegraphics[width=8.5cm]{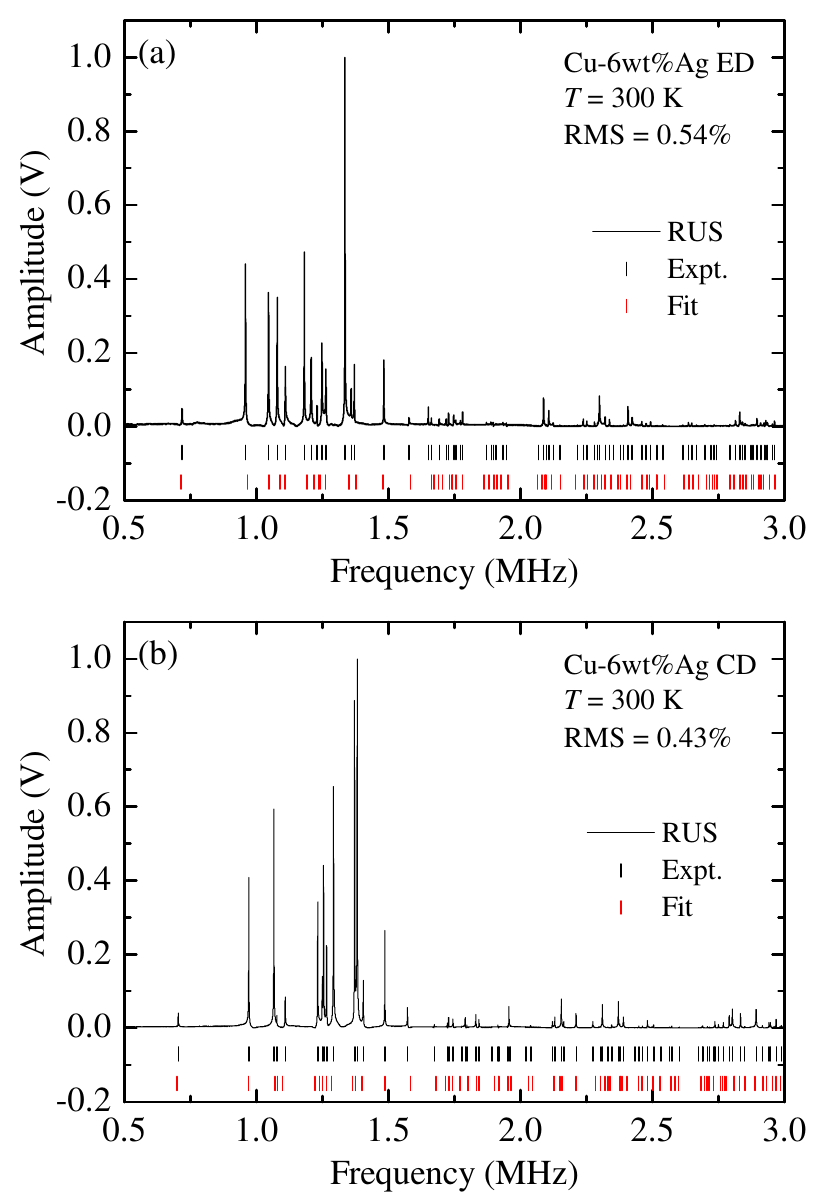}
\vspace{-10pt}
\caption{RUS spectra of Cu-6wt\%Ag alloy wire samples at room temperature. (a) ED; (b) CD. The black and red bars represent the experimental and calculated frequencies of resonant modes, respectively. }
\label{Fig2}
\end{figure}

Elastic properties of Cu-6wt\%Ag alloy wires were investigated by two different ultrasonic related techniques: RUS and ultrasound pulse echo. For RUS measurements, the Cu-6wt\%Ag samples were polished into parallelepipeds, and the dimensions were 0.957$\times$1.243$\times$1.319 mm$^{3}$ (ED) and 0.916$\times$1.219$\times$1.349 mm$^{3}$ (CD). RUS measurements were carried out in a lock-in technique by sweeping frequency from 0.3 to 3 MHz at stabilized temperatures ranging from 5.4 to 300 K in a helium-flow cryostat (OptistatCF, Oxford). A schematic of the RUS apparatus is presented in Fig.~\ref{Fig1}(b). The sample is corner-touch installed between two LiNbO$_3$ transducers with the bottom one acting as ultrasound transmitter (TX) and the top one as receiver (RX). When frequency of the excitation approaches a characteristic vibrational mode of the sample, a resonant peak will be picked up in RX. Alumina hemispheres were used between LiNbO$_3$ and sample, for insulation and mechanical protection \cite{ultrasounddevice2020}. For more details about RUS technique, we refer to Refs.~\cite{Migliori-RUS,RA2019RUStoolbook}.

For ultrasound pulse-echo measurements, the elastic constants were determined by measuring the ultrasound velocity in the principle of time-of-flight \cite{Luthi-Acoustics}. The Cu-6wt\%Ag samples were polished into parallelepipeds with dimensions 5.164$\times$3.929$\times$3.024 mm$^{3}$ (ED) and 5.972$\times$5.030$\times$2.733 mm$^{3}$ (CD). A pair of $Y$10-cut LiNbO$_3$-type transducers with fundamental vibration frequencies 30 MHz (longitudinal) and 18 MHz (transverse) were glued on the two parallel faces of each sample. The longitudinal ultrasound velocity $v_{L}$ with $\mathbf{k}\parallel\mathbf{u}$ (where $\mathbf{k}$ is the ultrasound propagation vector, and $\mathbf{u}$ is displacement vector) was measured by using a home-built integrated dual-mode pulse-echo ultrasonic measurement system at WHMFC \cite{Qiu-PulseEcho}. The measurement of transverse ultrasound velocity was also tried, but no detectable signal was observed, for unknown reasons. The elastic constant $C_{11}$ was calculated via $C_{11}=\rho v_{L}^2$, where $\rho = 8.732$ g/cm$^3$ is the mass density. These measurements were carried out under extreme conditions, with temperature down to 1.9 K, and magnetic field up to 50 T.

\section{\Rmnum{3}. Results and Discussion}

Figure ~\ref{Fig2}(a) and (b) respectively show representative vibrational spectra of Cu-6wt\%Ag alloy wire samples, taken at room temperature. A series of resonant peaks can be recognized. According to the principle of RUS technique, one can compute all the resonant frequencies with provided elastic constants $\{C_{ij}\}$ ($i,j=1...6$), sample dimensions, and density \cite{Migliori-RUS,Migliori-PhysicaB1993,Leisure-RUS}. Reversely, with a series of known resonant frequencies, density and sample dimensions, we can also derive $\{C_{ij}\}$ by a mathematical fitting process. The fitting  continues until the root mean square RMS $\equiv\sqrt{\{\sum_{i} [(f^i_{cal}-f_{exp}^i)/f^i_{exp}]^2\}/N}$ minimizes, where $N$ is the number of identified resonant peaks, and $f_{exp}^i$ and $f_{cal}^i$ are frequencies of the $i$th experimental (black bars in Fig.~\ref{Fig2}) and calculated (red bars) resonance modes, respectively. For isotropic polycrystalline samples, there are only two independent elements in the $\{C_{ij}\}$ tensor, viz $C_{11}$ and $C_{44}$, and all the elastic moduli can be calculated as follows \cite{Migliori-RUS}: \\
Shear modulus
\begin{equation}
    G=C_{44}=\frac{C_{11}-C_{12}}{2},
    \label{Eq1}
\end{equation}
Bulk modulus
\begin{equation}
    K=\frac{C_{11}+2C_{12}}{3},
    \label{Eq2}
\end{equation}
Young's modulus
\begin{equation}
    E=\frac{9KG}{3K+G},
    \label{Eq3}
\end{equation}
and Poisson's ratio
\begin{equation}
    \nu=\frac{3K-2G}{2(3K+G)}.
    \label{Eq4}
\end{equation}
At room temperature, the derived elastic constants for the ED and CD samples of Cu-6wt\%Ag are summarized in Tab.~\ref{Tab.1}. We notice that all elastic constants $C_{11}$, $K$, $E$ and $G$ are slightly larger in the ED sample, while their Poisson's ratios are comparable. Acoustic Debye temperature $\Theta_D$ and \textit{lattice} thermal conductivity $\kappa_L$ at 300 K ($T\gtrsim\Theta_D$) are also calculated according to the elastic constants \cite{Jia-LatticeThermalConduct,PanY-ScZrNbTaRhPd,Chen-U2012}. Both $\Theta_D$ and $\kappa_L$ are larger in the ED sample, as expected. However, since the total thermal conductivity should also include the contribution from electrons, it is premature to draw a precise conclusion before direct thermal conductivity measurements are made.

\begin{table*}[htbp]
\tabcolsep 0pt \caption{\label{Tab.1} Elastic constants of Cu-6wt\%Ag alloy wire samples in the absence of external magnetic field. $K$ - Bulk modulus, $E$ - Young's modulus, $G$ - Shear modulus, $\nu$ - Poisson's ratio. Acoustic Debye temperature $\Theta_D$ and lattice thermal conductivity $\kappa_L$ are calculated at 300 K according to the elastic constants \cite{Jia-LatticeThermalConduct,PanY-ScZrNbTaRhPd}. ``ED" and ``CD" represent the Cu-6wt\%Ag samples prepared by electroplastic drawing and conventional cold drawing, respectively.}
\vspace*{0pt}
\def\temptablewidth{1\textwidth}
{\rule{\temptablewidth}{1pt}}
\begin{tabular*}{\temptablewidth}{@{\extracolsep{\fill}}cccccccccc}
\hline
Material  &   $T$ (K)  &  Method & $C_{11}$ (GPa) & $K$ (GPa)   &    $E$ (GPa)  &   $G$ (GPa)     & $\nu$     &  $\Theta_D$  (K)    &  $\kappa_L$ (W/m$\cdot$K)
\\ \hline
\multirow{2}{*}{CD} & \multirow{2}{*}{300}    &  RUS  & 185.6 & 128.0  & 116.5  & 43.2   & 0.348    & 319.7   &  10.8
\\
 &    &  Pulse-echo  & 185.12  &  &   &    &     &    &
\\ \hline
\multirow{2}{*}{ED} & \multirow{2}{*}{300}    &  RUS  & 186.3  & 128.5 & 117.0  & 43.4   & 0.348    & 322.7   &   11.2
\\
 &    &  Pulse-echo  & 189.69  &  &   &    &     &    &
\\ \hline
\multirow{2}{*}{ED} & \multirow{2}{*}{78}    &  RUS  & 204.8 & 142.7  & 125.9  & 46.5   & 0.353    &    &
\\
 &    &  Pulse-echo  & 206.4  &  &   &    &     &    &
 \\ \hline
\multirow{2}{*}{ED} &  5.4    &  RUS  & 207.2 & 144.4  & 127.3  & 47.1   &  0.353   &    &
\\
 &  1.9  &  Pulse-echo  & 208.2 &  &   &    &     &    &
\\ \hline
\end{tabular*}
{\rule{\temptablewidth}{1pt}}
\end{table*}

\begin{figure*}[htbp]
\vspace{-0pt}
\hspace{0pt}
\includegraphics[width=17cm]{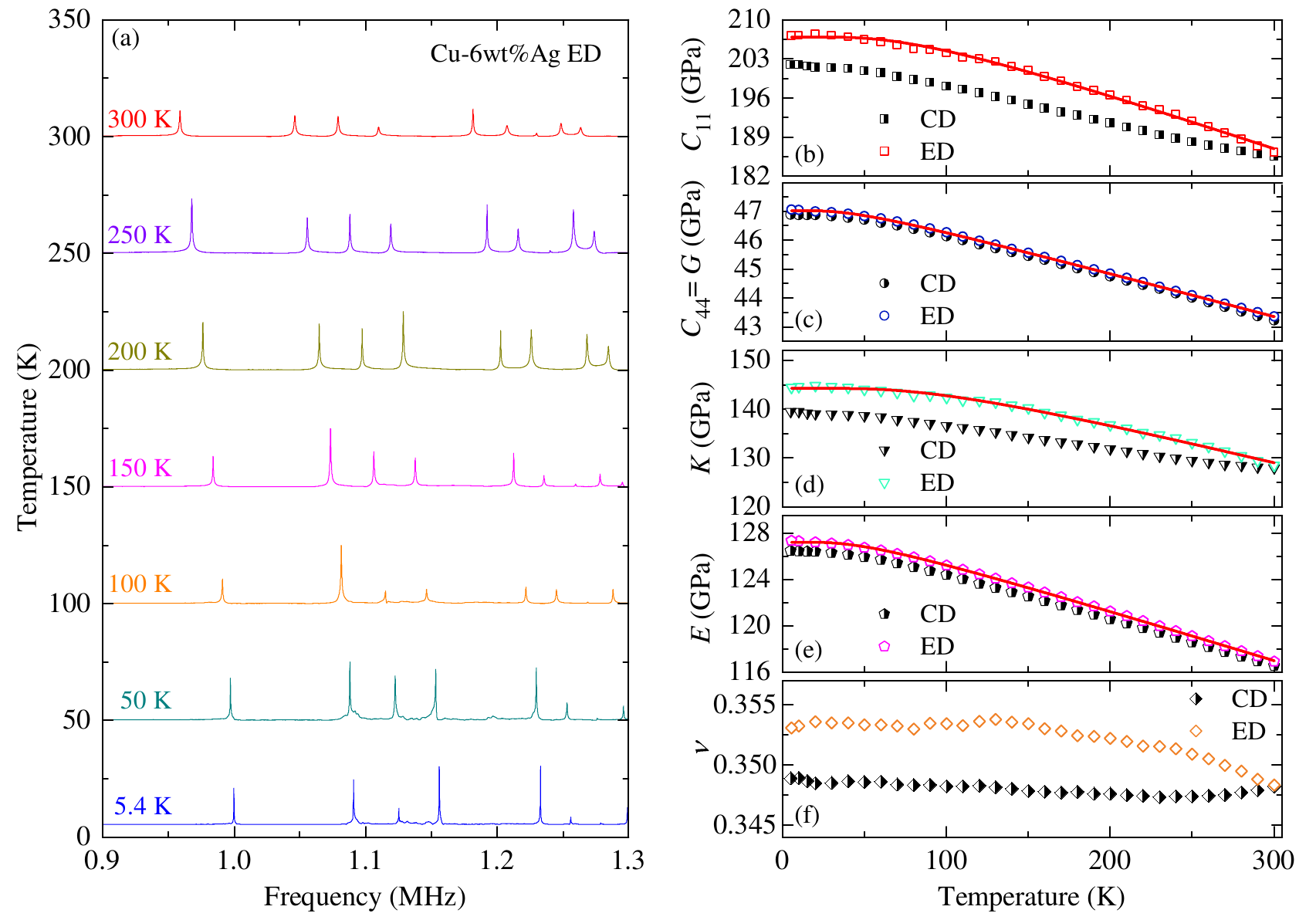}
\vspace{-0pt}
\caption{(a) RUS spectra of Cu-6wt\%Ag ED at selected temperatures in the range 5.4-300 K. For clarity, the spectra (0.9-1.3 MHz) are vertically offset. (b-f) temperature dependent $C_{11}$, $C_{44}(=G)$, $K$, $E$ and $\nu$ of the two measured samples. The red lines are fittings to Varshini's expression. }
\label{Fig3}
\end{figure*}

The elastic properties of Cu-6wt\%Ag samples at low temperature were also studied by RUS. Figure \ref{Fig3}(a) presents the spectra of the Cu-6wt\%Ag ED sample at selected temperatures between 5.4-300 K. One clearly finds that all the resonant peaks move to higher frequency upon cooling, implying that the material hardens at low temperature. The same trend is also seen in the CD sample (data not shown). The extracted temperature dependent elastic constants are displayed in Fig.~\ref{Fig3}(b-f). We find $C_{11}$, bulk modulus $K$, shear mudulus $G$, and Young's modulus $E$ all increase as temperature decreases, while for below $\sim 50$ K, these elastic constants tend to level off; the overall temperature dependencies can be well described by an empirical Varshni's expression $c(T)=c_0-s/(e^{T^*/T}-1)$, where $c = C_{ij}$, $K$, $G$, or $E$ \cite{varshni1970}, as denoted by the red lines. Another prominent feature is that, all through the temperature range 5.4-300 K, all these elastic constants are larger in the ED sample. In particular, at the liquid nitrogen temperature where Cu-6wt\%Ag alloy wires are used as current carrier of pulsed magnets, the electroplastic drawing treatment can enhance the moduli $C_{11}$, $K$, $G$ and $E$ by 2.7(1)\%, 4.3(1)\%, 0.3(1)\%, and 0.7(1)\%, respectively. Poisson's ratio $\nu$ is another physical quantity that should be discussed here. $\nu$ measures the extent of expansion of a material in directions perpendicular to the specific direction of compressing. At 300 K, $\nu=0.348$ for both samples. As is well known, $\nu$ sits in the range 0.25 - 0.35 for most conventional metals \cite{Greaves-Poisson_NM2011}; this is the case for both ED and CD Cu-6wt\%Ag alloys. However, the temperature dependent $\nu$ of the two samples behave differently: while $\nu$ of the CD sample is almost constant, data of the ED sample increase slightly at low temperature. Therefore, on the whole, $\nu$ is a little larger in the ED sample. Since $\nu$ characterizes the ductility, the relatively larger $\nu$ in ED implies more ease in coil winding, which is also advantageous for pulsed magnet application \cite{PanY-ScZrNbTaRhPd}.

\begin{figure}[htbp]
\vspace{-0pt}
\hspace{-0pt}
\includegraphics[width=8.5cm]{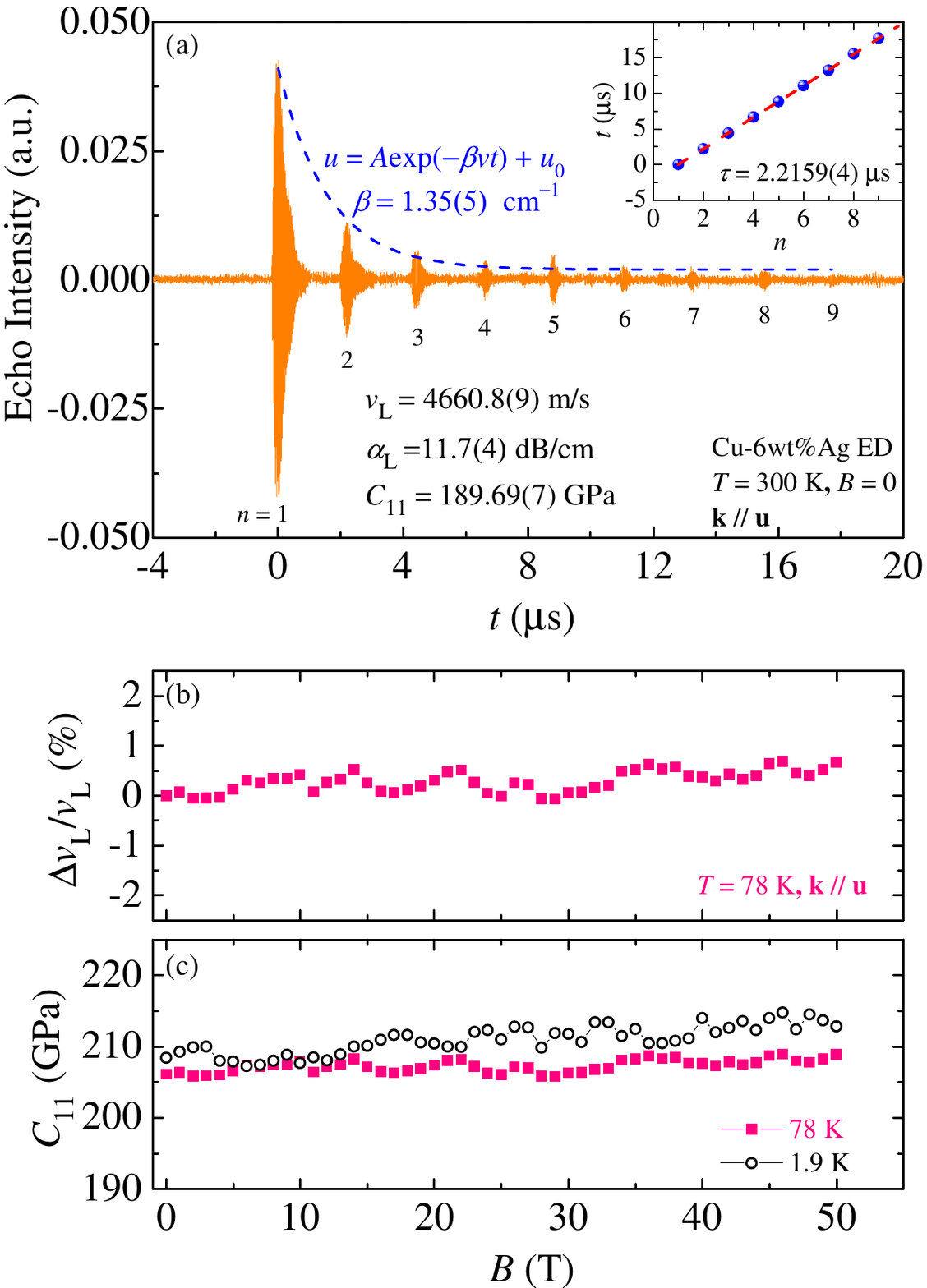}
\vspace{-15pt}
\caption{(a) Ultrasound pulse-echo measurements of Cu-6wt\%Ag ED sample at 300 K. The measurement was made with $\mathbf{k}\parallel\mathbf{u}$. The amplitude of the echos decays exponentially which yields the ultrasound attentuation coefficient $\alpha_L=11.7(4)$ dB/cm. The inset displays a plot of ultrasound propagation time vs the number of echo, the slope of which leads to the average time interval $\tau=2.2159(4)$ $\mu$s. (b) Field dependent $\Delta v_L/v_L$ at 78 K. (c) Field dependence of $C_{11}$ at 78 and 1.9 K.}
\label{Fig4}
\end{figure}

The elastic constants measured from RUS can be verified by an alternative ultrasound technique, the pulse-echo measurement. The principle of ultrasound pulse-echo is sketched in Fig.~\ref{Fig1}(c). To start with, representative result for the ED sample at 300 K is displayed in Fig.~\ref{Fig4}(a). After a single excitation of longitudinal ultrasound pulse ($\mathbf{k}\parallel\mathbf{u}$), a series of echos are clearly identified, whose amplitudes decay following an exponential law and can be well fit to $u(t)=A\exp{(-\beta vt)}+u_0$, where $A$ is a prefactor, and $u_0$ is a constant. The fitting yields $\beta = 1.35(5)$ cm$^{-1}$ which in turn gives rise to the ultrasound attenuation coefficient $\alpha_L=11.7(4)$ dB/cm. The position of each echo is plotted vs. $n$ in the inset to Fig.~\ref{Fig4}, the slope of which leads to the average time interval between echos, $\tau=2.2159(4)$ $\mu$s. The longitudinal ultrasound velocity can thus be calculated, $v_L=4660.8(9)\times 10^3$ m/s. With these, $C_{11}=189.69(7)$ GPa can be deduced. This result is in close agreement with that determined by RUS, cf. Tab. \ref{Tab.1}. The CD sample was also measured by ultrasound pulse echo experiment, and $C_{11}=185.12(8)$ GPa was obtained. It is important to mention that both RUS and pulse echo measurements reveal enhanced $C_{11}$ in the ED sample, implying that electroplastic drawing procedure does improve the elastic performance of Cu-6wt\%Ag alloy wire, albeit that such an improvement is weak. A natural question, then, concerns why the electroplastic drawing procedure can lift the UTS substantially, but only enhances the elastic properties relatively weakly? In order to clarify this issue, more investigations based on micro-structure experiments and systematic stress-strain measurements will be needed in the future \cite{Chen-CuAgUnpublish}.

To explore the performance of the ED sample under extreme conditions, we also investigated its elastic properties by ultrasound pulse echo measurements at low temperature down to 1.9 K and high magnetic field up to 50 T. By the so-called quadrature procedure and the phase-shift analysis \cite{Wolf-PFPulseEcho}, the relative change in the ultrasound velocity $\Delta v_L/v_{L}$ is measured as a function of magnetic field at fixed temperatures. For example, $\Delta v_L/v_{L}$ at 78 K is shown in Fig.~\ref{Fig4}(b); furthermore, the converted $C_{11}$ is presented in Fig.~\ref{Fig4}(c). Irrespective of the relatively large noise, a slight increase of $C_{11}$ with $B$ is visible, for both 78 K and 1.9 K, while no drastic softening can be seen up to 50 T. Note that for pulsed magnet application, the Cu-Ag alloy wires are used when immersed in liquid nitrogen. In this sense, our results thus demonstrate that Cu-6wt\%Ag ED exhibits a good elastic performance for this purpose. Here we must admit that the measurement precision under pulsed field is much lower than in the absence of field, likely due to the sample vibration caused by pulsed field and/or Eddy-current heating effect. Technically how to reduce the noise level will be a special concern for us in the future.

\section{\Rmnum{4}. Conclusions}

In conclusion, by a combination of resonant ultrasound spectroscopy and ultrasound pulse-echo experiments, we systematically studied the elastic properties of Cu-6wt\%Ag alloy wires. Our results show that an electroplastic drawing procedure can improve the elastic performance of Cu-6wt\%Ag alloy wires, and the elastic performance remains excellent under extreme conditions, e.g., ultra-high magnetic field, nitrogen / helium cryogenic liquids.

\section{Acknowledgments}

The authors are grateful to Haibin Yu for helpful discussions, and Alexey Suslov for technical aids. This work is supported by National Key R\&D Program of China (2022YFA1602602, 2023YFA1609600), National Natural Science Foundation of China (U23A20580), the open research fund of Songshan Lake Materials Laboratory (2022SLABFN27), Beijing National Laboratory for Condensed Matter Physics (Grant No. 2024BNLCMPKF004), Guangdong Basic and Applied Basic Research Foundation (2022B1515120020), and interdisciplinary program of Wuhan National High Magnetic Field Center at Huazhong University of Science and Technology (WHMFC202132).


\providecommand{\newblock}{}

\end{document}